\documentstyle[11pt,IAU207_pasp,twoside,psfig]{article}
\def\edcomment#1{\iffalse\marginpar{\raggedright\sl#1\/}\else\relax\fi}
\reversemarginpar

\begin{document}
\title{Globular Clusters around Galaxies in Groups}
 \author{Cristiano Da Rocha \& Claudia Mendes de Oliveira}
\affil{Instituto de Astronomia, Geof\'{\i}sica e Ci\^encias Atmosf\'ericas,
Universidade de S\~ao Paulo, Av. Miguel Stefano 4200, 04301-904, S\~ao Paulo --
SP, Brazil}
\author{Michael Bolte}
\affil{UCO/Lick Observatory, Department of Astronomy and Astrophysics,
University of California, Santa Cruz, California 95064}
\author{Bodo L. Ziegler}
\affil{Universit\"atssternwarte G\"ottingen, Geismarlandstr. 11, 37083
G\"ottingen, Germany}
\author{Thomas H. Puzia}
\affil{Universit\"atssternwarte M\"unchen, Scheinerstr. 1, D--81679
M\"unchen, Germany}

\begin{abstract}

We have obtained deep photometry of NGC 1199 (in HCG 22) and NGC 6868 (in
the Telescopium group). Both galaxies are the optically brightest galaxies
of their groups. Our analysis of $B$ and $R$ images taken with the Keck
II and the VLT/ESO telescopes, detected a population of globular clusters
around both galaxies, with total specific frequencies $S_N=1.7\pm0.6$
for NGC 1199 and $S_N = 1.3\pm0.6$ for NGC 6868. The color distributions
of the globular cluster systems shows bimodal peaks centered at $(B-R)_0
= 1.13\pm0.10$ and $1.42\pm0.10$ (NGC 1199) and $(B-R)_0=1.12\pm0.10$
and $1.42\pm0.10$ (NGC 6868).  
\end{abstract}

\section{Introduction}

The properties of extragalactic globular cluster systems (GCS) --
specific frequency, color distribution and radial profile shape -- may
be expected to vary as a function of the environment of the host galaxy.

The main goal of this work is to determine the properties of the globular
cluster systems of two galaxies in small groups: the central elliptical
galaxy in HCG 22 (NGC 1199) and the suspected merger in the center of the
Telescopium group (NGC 6868) (Hansen, J\o rgensen \& N\o rgaard-Nielsen,
1991).

We assume distance moduli of $(m-M)_V = 32.6$ and $32.1$ for NGC 1199 and
NGC 6868 respectively as determined from surface-brightness fluctuations
Tonry {\em et al.} (2001).

\section{Observations and Data Reduction}

The images of HCG 22 were obtained with the Keck II telescope, using
LRIS, in $B$ and $R$ with total exposure times of 720 and 630 seconds
and average seeing of 0.77 and 0.74 arcsec respectively. The NGC 6868
images were obtained with the VLT/ESO, using FORS1, in $B$ and $R$ with
total exposure times of 900 and 810 seconds and average seeing of 0.76
and 0.73 arcsec respectively. Images in $B$ and $R$ of similar depth were
obtained centered on a position 10$^{\prime}$ from the galaxy center to
be used as control field for background subtraction.

The light profile of the bright galaxies was modeled with ELLIPSE/STSDAS
and we used for SExtractor (Bertin \& Arnouts, 1996) for detection and
photometry of the faint objects.

\section{Specific Frequency and Radial Profile Modeling}

In order to calculate the specific frequency, the observed number counts
were corrected by the lost area due to masks and unobserved regions. We
fit a Gaussian ($M_{V} = -7.6$ and $\sigma = 1.18$, Drenkhahn \& Richtler,
1999), to the bright, measured GCLF and extrapolate the number of GC's
over all magnitudes in our systems.  Those counts were used as boundary
conditions to integrate the radial profile models and we obtained the
total number of objects in the GCS's.

We tried two kinds of radial profiles, a ``core model'' profile
($\rho=\rho_0(r_c^{\alpha} + r^{\alpha})^{-1}$) with the core radius
proposed by Forbes {\em et al.} (1996) ($r_c(kpc)=-(0.62\pm0.1)\cdot
M_V-11.0$) and a power--law profile ($\rho \propto r^{\alpha}$).

For NGC 1199 the slope of the power--law profile is --2.4.  This slope
is very steep, leading to unrealistic values.  Using this slope for the
outer region of the ``core model'' we estimated a $S_N = 1.7\pm0.6$,
which is smaller than the mean value found for a typical small group
elliptical (Harris, 1991, gives $2.6\pm0.5$), but still in agreement
within the errorbars.  For NGC 6868 the slope of the power--law is --1.4.
We estimated $S_N = 1.4\pm0.6$ with this profile. The ``core model'',
with the same slope, gives us $S_N = 1.3\pm0.6$. Those estimates are
a factor of two smaller than typical indicating a poor system around
this galaxy. The radial profiles for both galaxies and the best--fit
power--law and ``core model'' are shown in figure 1.

\section{Color Distribution}

The KMM test (Ashman, Bird \& Zepf, 1994) was used to estimate the
parameters of the color distributions. We have detected bimodal
populations for both galaxies with 100\% confidence level. The peaks
are located at $(B-R)_{0}=1.13\pm0.10$ and $1.42\pm0.10$ for NGC 1199
and $(B-R)_{0}=1.12\pm0.10$ and $1.42\pm0.10$ for NGC 6868. Using Reed,
Harris \& Harris (1994) color--metallicity relation, we have $[Fe/H] =
-1.45\pm0.23~dex$ and $-0.55\pm0.23~dex$ for the blue and red peaks of NGC
1199, respectively and $[Fe/H] = -1.48\pm0.23~dex$ and $-0.55\pm0.23~dex$
for the blue and red peaks of NGC 6868, respectively.

\acknowledgments

This work is supported by FAPESP PhD. grant No. 96/08986--5 and Pronex NExGAL.

\begin{figure}[h]
\centerline{\vbox{
\psfig{figure=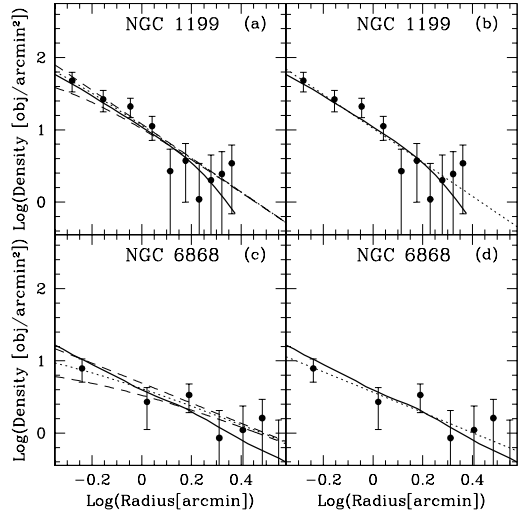,width=9.6cm,angle=0}
}}
\caption{Radial profiles of globular cluster candidates. Panel (a)
shows the radial profile of GC's around NGC 1199 with the ``core model''
profile (dotted line) and its upper and lower limits (dashed lines)
and the galaxy light in the B band (continuous line) overplotted.
Panel (b) shows the the power--law profile (dotted line) overplotted
for NGC 1199. Panels (c) and (d) show the same information of panels
(a) and (b), respectively, for the GCS around NGC 6868.}
\end{figure}

\end{document}